\documentstyle[11pt]{article}
\begin{document}

\title{The problem of quantum chaotic scattering with direct processes reduced to the one without}

\author{$^1$V\'{\i}ctor A.Gopar and $^2$Pier A. Mello}
\maketitle
$^1$Instituto de F\'{\i}sica, Universidad Nacional 
Aut\'onoma de M\'exico, 01000 M\'exico D.F., M\'exico

$^2$D\'epartement de Physique Th\'eorique, Universit\'e de Gen\`eve, CH-1211 Gen\`eve 4, Switzerland

\begin{abstract}
We show that the study of the statistical properties of the scattering matrix
$S$ for quantum chaotic scattering in the presence of direct processes 
(characterized by $\overline S \neq 0$, $\overline S$ being the average $S$
matrix) can be reduced to the simpler case where direct processes
are absent ($\overline S = 0$).
Our result is verified with a numerical simulation 
of the two-energy autocorrelation for two-dimensional $S$ matrices.
It is also used to extend Wigner's time delay 
distribution for one-dimensional $S$ matrices, recently found for
$\overline S = 0$, to the case $\overline S \neq 0$; this extension is 
verified numerically. As a consequence of our result, future calculations
can be restricted to the simpler case of no direct processes.
\end{abstract}

The problem of chaotic wave scattering is of great interest in nuclear,
molecular and mesoscopic physics, in optics and in the microwave domain.
Common features have been found for system sizes spanning such a large
range because of the generality of the phenomena involved.

In quantum-mechanical scattering problems with a chaotic classical dynamics 
one aims at studying the statistical properties of the scattering
matrix $S$. Recently, this study has been further 
motivated by experiments on quantum electronic transport in 
mesoscopic systems \cite{Huibers}.

The {\it one-energy} statistical distribution of the $S$ matrix has been
described successfully by an {\it information-theoretic} model that
incorporates precisely the {\it physical information}
that is relevant for a wide class of systems \cite
{hua,mello-pereyra-seligman,baranger-melloEPL}. That information specifies:
{\bf 1)} General properties: 
{\it i)} {\it flux conservation} (giving rise to unitarity of the 
$S$ matrix), 
{\it ii)} {\it causality} and the related analytical properties of $S(E)$ in
the complex-energy plane, and
{\it iii) }the presence or
absence of time-reversal and spin-rotation symmetry, that determines the 
{\it universality class}: orthogonal, unitary or symplectic 
(designated as $\beta =1,2,4$) \cite{mehta,porter} and restricts further the 
structure of $S$: unitary symmetric, unitary or unitary self-dual, 
respectively.
{\bf 2)} A specific property: the ensemble average 
$\left\langle S\right\rangle$, identified with the 
energy average $\overline{S}$, also known as the {\it optical }$S$
{\it matrix} 
\cite{feshbach}, which controls the presence of {\it prompt}, or
{\it direct processes} in the scattering problem. In this procedure one 
constructs the statistical distribution of
$S$ using only the above physical information 
--expressible entirely in terms of $S$ itself-- without ever invoking any
statistical assumption for the underlying Hamiltonian, that never enters 
the analysis.
The resulting $S$-matrix distribution, known as {\it Poisson's kernel}, 
reproduces well the statistical scattering properties of
ballistic cavities with a chaotic classical dynamics \cite
{baranger-melloEPL}.

The joint statistical distribution of the $S$ matrix at {\it two or more
energies} has escaped, so far, an analysis within the philosophy described
above (some aspects of the two-point problem have
been studied assuming an underlying Hamiltonian described by a Gaussian
ensemble, as in Refs. \cite{lewenkopf,fyodorov}). An approach
coming close to that philosophy was initiated in Ref. \cite
{gopar-mello-buettiker}, with a study of the simplest quantity of a
two-point character: the statistical distribution of the time delay --that
involves the energy derivative of the $S$ matrix-- arising in the scattering
process \cite{bauer-mello-mcvoy}. The physical motivation of
Ref. \cite{gopar-mello-buettiker} was
the study of the electrochemical capacitance of a mesoscopic system \cite{buettiker93/93/93/94}. 
The analysis was done for a cavity attached to one lead that can
support only one open channel (so that $S=e^{i\theta }$is a $1\times 1$
matrix), for arbitrary $\beta $ and $\overline{S}=0$. It was based on a
conjecture by Wigner \cite{wigner}, that assumes invariance of the 
statistics
of the poles and residues of the related $K$ matrix under the
same transformation $S\rightarrow e^{i\phi /2}Se^{i\phi /2}$ that defines
the invariant one-point measure \cite{dyson} which, in the one-channel
case, is $d\theta /2\pi $. That study was generalized to an arbitrary number
of channels $N$ in Refs. \cite{brouwer-frahm-beenakker,piet}, again for 
$\overline{S}=0$: Wigner's conjecture was formulated and proved as the
invariance of the $k$-point probability distribution of the $N$-channel $S$
matrix under the transformation that defines the invariant measure for the
universality class $\beta $. 

Of physical importance is the case $\overline{S}\neq 0$: it 
corresponds to situations where direct processes are not negligible. 
For example, the case
$\overline{S}=0$ describes chaotic cavities with ballistic 
point contacts, while the coupling to leads containing tunnel barriers 
produces direct reflection and thus $\overline{S}\neq 0$ 
\cite{baranger-melloEPL,piet1}. A more complex combination of direct 
processes is described in Ref. \cite{baranger-melloEPL}.

In the present article we find a transformation that
relates the $k$-point distribution of the $N$-channel $S$ matrix for the
case $\overline{S}\neq 0$ to that for $\overline{S}=0$, thus relating the
problems in the presence and in the absence of direct processes. 
In Ref. \cite{Engelbrecht-HRTW-Nishioka}, the problem with a nondiagonal
$\overline S$ was reduced to that with a diagonal and real $\overline{S}$.
Here we show that the problem can be reduced further
to one characterized by $\overline S = 0$.
That transformation, used in the past in relation with one-point 
functions \cite{friedman-mello}, contains only the $S$ matrix and its 
average
$\overline{S}$. It is used here to extend to the case $\overline{S}\neq 0$ 
the time-delay distribution of Ref. \cite
{gopar-mello-buettiker} and is verified in this case by comparing our
results with some known ones and with numerical simulations.
It is also verified numerically for the two-point correlation function of 
two-dimensional $S$ matrices.
We believe that our result is appealing because of its conceptual simplicity 
and may open the way for further studies in this field.

We first summarize the information-theoretic model of Refs. 
\cite{hua,mello-pereyra-seligman,baranger-melloEPL}. The starting point is
the measure $d\mu ^{(\beta )}(S)$ for $n\times n$ $S$ matrices, that
remains {\it invariant} under the symmetry operation for the universality
class $\beta $ \cite{mehta,porter,dyson}. The ensemble average of $S$
vanishes for the invariant measure, and so does the prompt component. An
ensemble for which $\left\langle S\right\rangle =\overline{S}\neq 0$
contains more information than the invariant-measure: it can be constructed
as $p_{\overline{S}}^{^{(\beta )}}(S)d\mu ^{(\beta )}(S)$  
and the information associated with it is defined as 
$\int p_{\overline{S}}^{^{(\beta )}}(S)\ln \left[ p_{\overline{S}}^{^{(\beta
)}}(S)\right] d\mu ^{(\beta )}(S)$.
We assume to be far from thresholds and recall that $S(E)$ is {\it analytic}
in the upper half of the complex-energy plane (causality) \cite{mcvoy}. 
The study of the statistical properties of $S$ is simplified by idealizing 
$S(E)$, for real $E$, as a {\it stationary
random function} satisfying the condition of {\it ergodicity}, which in turn
implies the equivalence of energy and ensemble averages \cite{yaglom}. We 
thus have the {\it analyticity-ergodicity} requirements (AE), that imply 
$\left\langle S_{ab}^pS_{cd}^q \cdots \right\rangle =\left\langle
S_{ab}\right\rangle ^p\left\langle S_{cd}\right\rangle ^q \cdots $, 
so that averages of products of $S$-matrix elements not involving complex
conjugation can be written in terms of the matrix $\left\langle S \right\rangle = 
\overline{S}$. The one-energy probability density 

\begin{equation}
\label{poisson}p_{\overline{S}}^{(\beta )}(S)=V_\beta ^{-1}\frac{[{\rm det}%
(I-\bar{S} \bar{S}^{\dagger })]^{(\beta n+2-\beta )/2}}{\mid {\rm %
det}(I-S \bar{S}^{\dagger })\mid ^{\beta n+2-\beta }}, 
\end{equation}
where $V_\beta $ is a normalization constant, is known as Poisson's kernel: 
it satisfies the AE
requirements \cite{hua} and the associated information is less than or equal
to that of any other probability density satisfying AE for the same $%
\overline{S}$ \cite{mello-pereyra-seligman}. In Eq. (\ref{poisson}), the
elements of $S$ are assumed to be complex numbers for $\beta =1,2$ and
quaternions for $\beta =4$; for the definition of the determinant of a
quaternion matrix, see Ref. \cite{mehta}, p. 126 .

Now consider the transformation  
\begin{equation}
\label{S0}S_0= t_1^{\prime -1}(S-r_1)(1-r_1^{\dagger}S)^{-1}
t_1^{\dagger }, 
\end{equation}
where $r_1$, $t_1$, $r_1^{\prime }$, $t_1^{\prime }$ are the $n\times n$
blocks of the matrix 
\begin{equation}
\label{S1}S_1=\left[ 
\begin{array}{cc}
r_1 & t_1^{\prime } \\ 
t_1 & r_1^{\prime } 
\end{array}
\right] , 
\end{equation}
which has the symmetry associated with the universality class $\beta $. One
can prove the following statement \cite{hua,piet,piet1,friedman-mello}: if
the one-energy distribution of $S$ is Poisson's measure (\ref{poisson}) and
we identify $r_1$  in the transformation (\ref{S0}) 
with $\overline{S}$, 
the one-energy distribution of $S_0$ is the invariant
measure $d\mu ^{(\beta )}(S_0)$. In other words, the transformation 
(\ref{S0}%
), with $r_1=\overline{S}$, transforms the problem with direct processes to
one without ($\overline{S_0}=0$), as far as the one-energy distribution 
goes.

Now suppose that {\it at every energy} $E$ we subject $S(E)$ to the
transformation (\ref{S0}), always with the same $\overline{S}$. We
prove below the following statement: {\it the joint statistical properties 
of
the transformed }$S_0(E_1)$, $S_0(E_2)$ {\it , etc. are precisely the
ones associated with the problem without direct processes,
characterized by $\overline{S_0}=0$}. In other words,
the above transformation relates $S(E)$, understood as
a {\it stationary random function of energy}, for a problem with direct 
processes to one without.

We prove the above statement for $S$ unitary and symmetric ($\beta =1$),
the proof for the other symmetries being similar.
Consider first the case of $\overline{S}$ diagonal and real.
$S$ can be written in terms of the real amplitudes 
$\gamma_{\lambda a}$'s and the energy levels $E_\lambda$'s, as  

\begin{equation}
\label{SE} S(E)=[1 + i K(E)][1-iK(E)]^{-1} ,
\end{equation} 
where the $K$ matrix is given by 

\begin{equation} 
\label{KE} K_{ab}(E)=\sum_{\lambda} \frac{\gamma_{ \lambda a}\gamma_{ 
\lambda b}}{E_{\lambda} - E}.
\end{equation}
The $\gamma _{\lambda a}$'s are uncorrelated Gaussian variables and the 
$E_{\lambda}$'s follow the statistics of the Gaussian Orthogonal Ensemble 
\cite{mehta,porter}, with average spacing $\Delta$. 
$\overline S(E)$ is given by \cite{Engelbrecht-HRTW-Nishioka}

\begin{equation}
\label{Sbar} \overline{S(E)}_{a b} = 
(1-y_{aa})(1+y_{bb})^{-1}\delta_{a b},
\end{equation}
where $y_{aa}=\pi\left\langle \gamma_{\lambda a} ^2\right\rangle /\Delta$.
Just as $S$ and $K$ are related by Eq. (\ref{SE}), the transformed $S_0$, Eq. (\ref{S0}), can be written in terms of a matrix $K_0$ given by

\begin{equation}
\label{K0} K_0 =i (1-S_0)(1+S_0)^{-1}.
\end{equation}
Substituting the transformation (\ref{S0}) into (\ref{K0}), 
with $S$ given by (\ref{SE}), we finally find

\begin{equation}
          (K_0)_{ab}= y_{aa}^{-1/2} K_{ab}y_{bb}^{-1/2}.
\end{equation}

Thus, the $\gamma_{\lambda a}$'s are rescaled by a constant
($\gamma_{\lambda a} \rightarrow y_{aa}^{-1/2}\gamma_{\lambda a}$), which
is the appropriate one to ensure  $\overline{S_0(E)}=0$.
Therefore, under the transformation (\ref{S0}), the 
$\gamma_{\lambda a}$'s retain their Gaussian distribution, 
while the energy levels $E_{\lambda}$'s are unchanged.
We conclude that the statistical properties of the transformed random function $S_0(E)$ 
are the same as those of a statistical $S(E)$ matrix without
direct processes. This proves our statement for $\overline{S}$ 
diagonal and real. The case of arbitrary $\overline{S}$ can now be reduced 
to the above one using the results of Ref. \cite{Engelbrecht-HRTW-Nishioka}.

As a first application of the above statement, we extend to
$\overline{S}\neq 0$ the time-delay distribution found in
Ref. \cite{gopar-mello-buettiker} for one spatial channel, 
$\overline{S}=0$ and arbitrary $\beta $.
We consider Eq. (\ref{S0}) for $n=1$, with $r_1=\overline{S}$;
omitting the unimportant phase factor $t_1^{*}/t_1$, we have 
\begin{equation}
\label{S0N1}S_0(S)=(S-\overline{S})(1-\overline{S}^{*}S)^{-1}, 
\end{equation}
where $S_0$ and $S$ can be written as $S_0=\exp (i\theta _0)$, 
$S=\exp (i\theta )$.

We are interested in the time delay \cite
{bauer-mello-mcvoy} $\theta ^{\prime }=d\theta /dE$; we express it in terms of $\theta _0$ and $\theta _0^{\prime }=d\theta _0/dE$ as 
\begin{equation}
\label{dtheta/dE}
\theta ^{\prime }=d\theta /dE =\left[1-\left| \overline{S}\right| ^2\right]
\left| 1+ \overline{S}^{*}e^{i\theta _0}\right| ^{-2}
d\theta _0/dE
\equiv f(\theta _0)\theta _0^{\prime }. 
\end{equation}
To find the distribution of $\theta ^{\prime }$ we need the
joint distribution of $\theta _0$ and $\theta _0^{\prime }$, 
$p_0^{(\beta )}(\theta _0,\theta _0^{^{\prime }})$, which, being the one for no 
direct processes, factorizes as \cite{brouwer-frahm-beenakker, piet}
\begin{equation}
\label{p(theta0 theta0')}p_0^{(\beta )}(\theta _0,\theta _0^{^{\prime}})
= p_0^{(\beta )}(\theta _0^{\prime })/2\pi. 
\end{equation}
It is convenient to express Eqs. (\ref{dtheta/dE}) and (\ref{p(theta0 theta0')}) in terms of the variables $u$, $u_0$ defined as
\begin{equation}
\label{u-u0}
u= 2\pi /\theta ^{\prime }\Delta , \;\;\; 
u_0=2\pi /\theta^{\prime }_0\Delta _0 , 
\end{equation}
where $\Delta$ and $\Delta_0$ denote the average level spacing for the 
problems described by $S$ and $S_0$, respectively. The above proof shows that
$\Delta = \Delta_0$, since the energy levels $E_\lambda$ are unchanged by the transformation 
(\ref{S0}). Thus, using (\ref{u-u0}) and $u_0 = 2\pi/{\theta_0}^{\prime}\Delta$, Eq. (\ref{dtheta/dE}) becomes
\begin{equation}
\label{u(u0)}u=u_0/f(\theta _0). 
\end{equation}
The joint distribution of the statistically independent variables $\theta_0, u_0$ [see Eq. (\ref{p(theta0 theta0')})], needed to find the distribution of $u$, is given by
\begin{equation}
\label{p(theta0 u0)}P_0^{(\beta )}(\theta _0,u_0)=
P_0^{(\beta )}(u_0)/2\pi , 
\end{equation}
where $P_0^{(\beta )}(u_0)$ was obtained in Eqs. (16), (17) of  Ref. \cite{gopar-mello-buettiker} [and denoted there by $\widehat{P}(u)$] as 

\begin{equation}
\label{p(u0)}P_0^{(\beta )}(u_0)=\frac{\left( \beta /2\right) ^{\beta /2}}{%
\Gamma \left( \beta /2\right) }u_0^{\beta /2}e^{-\left( \beta /2\right)
u_0}. 
\end{equation}
From (\ref{u(u0)}), (\ref{p(theta0 u0)}), (\ref{p(u0)}) we write the 
distribution of $u$, $P_{\overline{_S}}^{(\beta )}(u)=
\left\langle \delta \left[ u - u_0 /f(\theta _0)\right] \right\rangle $, as

\begin{equation}
P_{\overline{_S}}^{(\beta )}(u)=\frac 1{2\pi }\int f(\theta_0)P_{0}^{(\beta 
)}(f(\theta _0)u)d\theta _0 
=\frac{\left( \beta /2\right)
^{\beta /2}}{2\pi \Gamma \left( \beta /2\right) }u^{\beta /2}\int_0^{2\pi
}\left[ f(\theta _0)\right] ^{1+\beta /2}e^{-\frac \beta 2uf(\theta
_0)}d\theta _0. 
\end{equation}
For the dimensionless time delay $\tau =1/u=\theta ^{\prime }\Delta/2\pi$ 
we finally find the probability density

\begin{equation}
\label{w(tau)}
w_{\overline{_S}}^{(\beta )}(\tau )
=\frac
{\left(\frac{\beta}{2}\right)^{\beta /2}}
{2\pi \Gamma \left(\frac{\beta}{2}\right)\tau^{2+\frac{\beta}{2} }}
\int_0^{2\pi }\left[ f(\theta _0)\right] ^{1+\frac{\beta}{2}}
e^{-\frac{\beta}{2\tau}f(\theta _0)}d\theta _0. 
\end{equation}

The calculation of $w_{\overline{S}}^{(\beta )}(\tau )$ 
for arbitrary $\overline{S}$ and $\beta $ is thus reduced to
quadratures [$f(\theta _0)$ is known, Eq. (\ref{dtheta/dE})]  
and the result coincides with our previous one 
$\cite{gopar-mello-buettiker}$ for $\overline{S}=0$.

Eq. (\ref{w(tau)}) is compared in Fig. 1 (a) with 
a numerical simulation that generates an ensemble
of $S$'s from resonances sampled from an unfolded Gaussian ensemble
for $\beta = 1$, the coupling amplitudes to the channel being independent 
Gaussian variables whose variance ensures $\overline{S}=1/2$.
The agreement is very good. 
Such a comparison was also made for $\beta =2,4$, although it is not 
illustrated here. For $\beta =2 $, Ref. \cite{fyodorov} gives 
$w_{\overline{_S}}^{(\beta )}(\tau )$ in analytical form in terms of Bessel
functions: we could not prove analytically the
equivalence with our result; however, numerically the two are indistinguishable.

As a further verification of our statement, consider the autocorrelation
function 
$c_{\overline{S}}(E)=\left\langle S^{*}_{11}(0)S_{11}(E)\right\rangle _{%
\overline{S}} - \left\langle S^{*}_{11}(0)\right\rangle _{\overline{S}}
\left\langle S_{11}(E)\right\rangle _ {\overline{S}}$. 
The quantity $\left| c_{\overline{S}}(E)\right|^2 $ was calculated by 
generating numerically ensembles of  
$2\times 2$ $S$ matrices, with $\overline{S}=0$ and $(1/2)I$,
$I$ being the unit matrix.
To the data for $\overline{S}=(1/2)I$ the transformation (\ref{S0}) 
was applied:
the results shown in Fig. 1 (b) are seen to be consistent with 
$\left| c_{\overline{S}=0}(E)\right|^2$.

Summarizing, we have found a transformation that relates the $S$ matrix 
$S(E)$, understood as a stationary random function of energy, for a problem 
with direct processes ($\overline{S}\neq 0$) to one without 
($\overline{S}=0$). 
An application was made to extend Wigner's time-delay 
distribution for one channel $S$-matrices, 
from $\overline S = 0$ to $\overline S \neq 0 $. 
A number of numerical simulations was made as a verification
of our transformation. 
Our result implies that future work on the
statistical properties of the $S$ matrix can be restricted to the simpler
case $\overline{S}=0$, and extended to the case 
$\overline{S}\neq 0$ $-$corresponding to more complex scattering systems with direct processes$-$ 
using the procedure described in this paper.

Part of this work was supported by DGAPA, DGSCA and CONACyT, M\'exico, and by the
Swiss National Science foundation. P.A.M. is
grateful to the \'Ecole de Physique de l' Universit\'e de Gen\`eve for its
hospitality during the time most of this work was accomplished. 

E-mail: gopar@fenix.ifisicacu.unam.mx, mello@fenix.ifisicacu.unam.mx

\begin{figure}[bp]
\caption{(a) The distribution of time delays for one channel 
and in the presence of direct processes ($\overline{S}=1/2$), for $\beta=1$. 
The crosses are proportional to the theoretical probability density of Eq. 
(\ref{w(tau)}), that was integrated numerically. The points with the finite-
sample error bar are the results of the numerical simulation 
described in the text. The agreement is excellent.
(b) The square of the autocorrelation function of the $S_{11}$ 
element of a two-channel $S$ matrix for $\beta =2$ as a function of the 
energy separation, obtained from a numerical simulation. The open circles 
and diamonds correspond to $\overline{S}=0$ and $(1/2)I$, respectively. The 
squares are the result of applying the transformation (\ref{S0}) to the data 
for $\overline{S}=(1/2)I$: they are seen to be consistent with the 
correlation 
for $\overline{S}=0$.}
\label{fig1}
\end{figure}

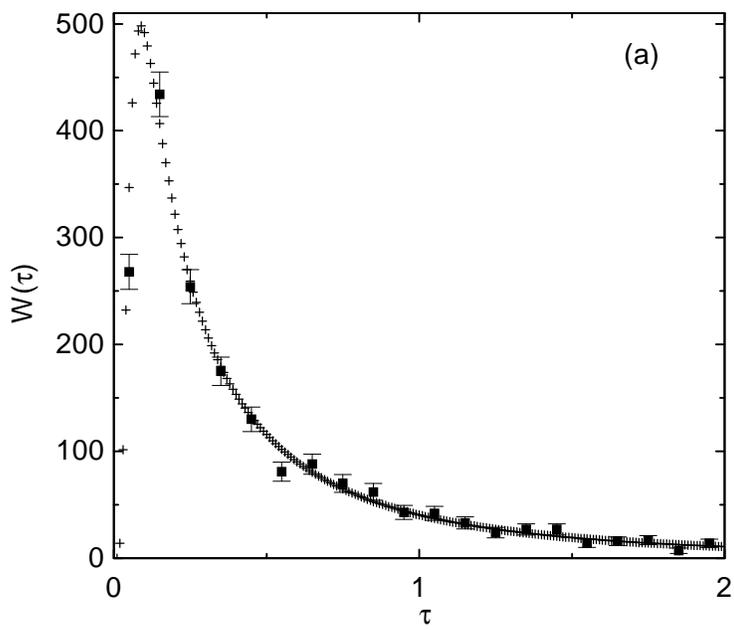
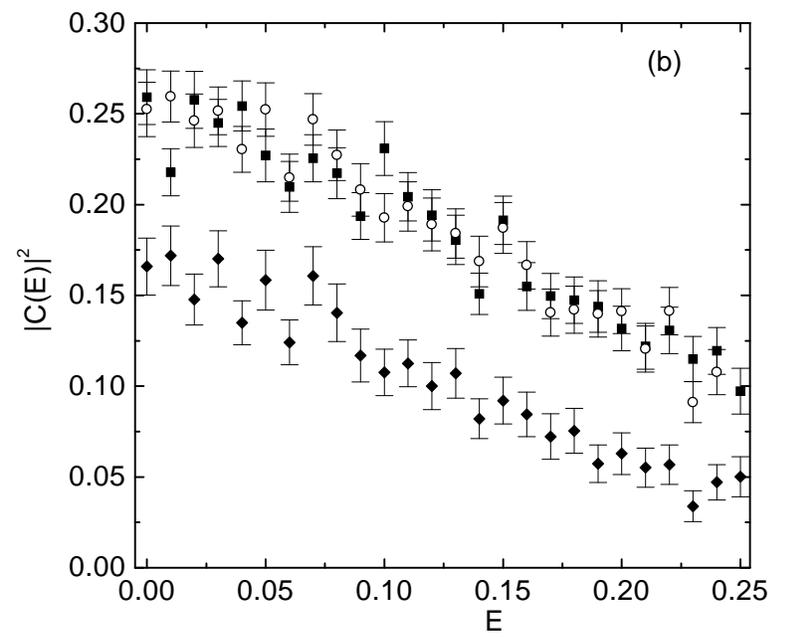
\end{document}